\documentclass[journal]{IEEEtran}
\usepackage{graphicx}
\usepackage{caption}
\usepackage{subcaption}
\usepackage{tikz}

\newcommand{\fig}[1]{Fig.~\ref{#1}}

\begin{document}
\title{Optimisation of the Read-out Electronics of Muon Drift-Tube Chambers for Very High Background Rates at HL-LHC and Future Colliders}

\author{\underline{S.~Nowak$^*$}\thanks{$^*$Corresponding author: nowak@mpp.mpg.de}, S. Abovyan, P. Gadow, K. Ecker, D. Fink, M. Fras, O.~Kortner, H.~Kroha, F.~M\"uller, R.~Richter, C.~Schmid, K.~Schmidt-Sommerfeld, Y. Zhao \\ \textit{Max-Planck-Institut f\"ur Physik, Munich}}

\maketitle
\pagestyle{empty}
\thispagestyle{empty}

\begin{abstract}
	In the ATLAS Muon Spectrometer, Monitored Drift Tube (MDT) chambers and sMDT chambers with half of the tube diameter of the MDTs are used for precision muon track reconstruction.
	The sMDT chambers are designed for operation at high counting rates due to neutron and gamma background irradiation expected for the HL-LHC and future hadron colliders.
	The existing MDT read-out electronics uses bipolar signal shaping which causes an undershoot of opposite polarity and same charge after a signal pulse.
	At high counting rates and short electronics dead time used for the sMDTs, signal pulses pile up on the undershoot of preceding background pulses leading to a reduction of the signal amplitude and a jitter in the drift time measurement and, therefore, to a degradation of drift tube efficiency and spatial resolution.
	In order to further increase the rate capability of sMDT tubes, baseline restoration can be used in the read-out electronics to suppress the pile-up effects.
	A discrete bipolar shaping circuit with baseline restoration has been developed and used for reading out sMDT tubes under irradiation with a 24~MBq $^{90}$Sr source.
	The measurements results show a substantial improvement of the performance of the sMDT tubes at high counting rates.
\end{abstract}

\section{Introduction}

The ATLAS Monitored Drift Tube (MDT) chambers~\cite{ATLAS} account for the vast majority of precision tracking chambers in the Muon Spectrometer of the ATLAS experiment at the Large Hadron Collider (LHC), where they have to cope with unprecedented radiation background of photons and neutrons in the energy range around 1~MeV.
With the upgrade of the LHC to the High Luminosity LHC (HL-LHC), which implies an increase of the radiation background rate by a factor of five, the rate capability of both the MDT detectors and their front-end electronics will exceed.
Therefore, new detectors, so-called small-diameter Muon Drift Tube (sMDT) chambers with 15~mm diameter, half of the MDT, have been developed.
These chambers, which are a strong candidate for precision muon detectors at future hadron colliders, are fully compatible with the MDT chambers in terms of read-out and services.
The sMDT tubes have an about one order of magnitude higher rate capability than the MDTs~\cite{sMDT}.
They can be operated at much shorter electronics dead time of about 200~ns and are much less sensitive to space charge effects.
The improved rate-capability of the sMDT chambers can, however, not be fully exploited with the present MDT front-end electronics due to limitations in the analog signal processing.

\section{Limitations of the Present Read-Out Electronics}

\begin{figure}[h]
	\centering
	\includegraphics[width=0.5\textwidth]{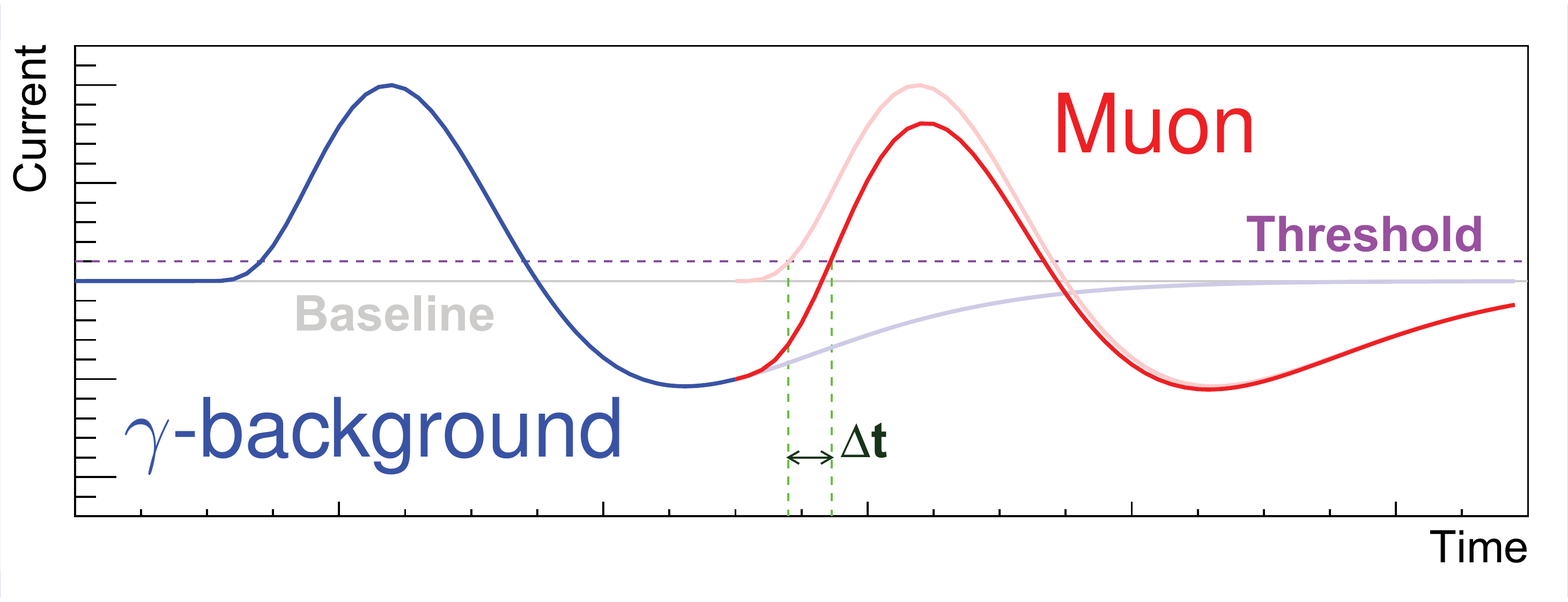}
	\begin{tikzpicture} [overlay]
		\node[text width=15cm, color=red] at (0.5\textwidth,2.8) {Muon};
		\node[text width=21.5cm, color=blue] at (0.5\textwidth,2.8) {$\gamma$-background};
	\end{tikzpicture}
	\vspace{-20mm}
	\caption{Illustration of signal pile-up effects with bipolar shaping~\cite{Pisa}. Due to the overlapping signal undershoot of the preceding pulse, the threshold crossing time of the successive pulse is shifted by $\Delta t$ and the amplitude is reduced.}
	\label{fig::pile_up}
\end{figure}

The present read-out chip used for both MDT and sMDT chambers is the so-called ASD (Amplifier, Shaper, Discriminator)~\cite{ASD}.
It uses a bipolar shaping scheme, which has the advantage of guaranteeing  baseline stability at high counting rates, but introduces an undershoot below the baseline with the same charge successive after a input pulse.
At high counting rates with the short programmable dead time of 200~ns of the ASD used for the sMDTs, signal pulses can overlap with the undershoot of preceding background pulses leading to a reduction of the signal amplitude and a delayed threshold crossing time and, therefore, to a degradation of efficiency and spatial resolution of the drift-tube.
This so-called pile-up effect is illustrated in \fig{fig::pile_up}.

\vspace{-10mm}

\begin{figure}[h!]
	\centering
	\begin{subfigure}[b]{0.4\textwidth}
		\centering
		\includegraphics[width=1.2\textwidth]{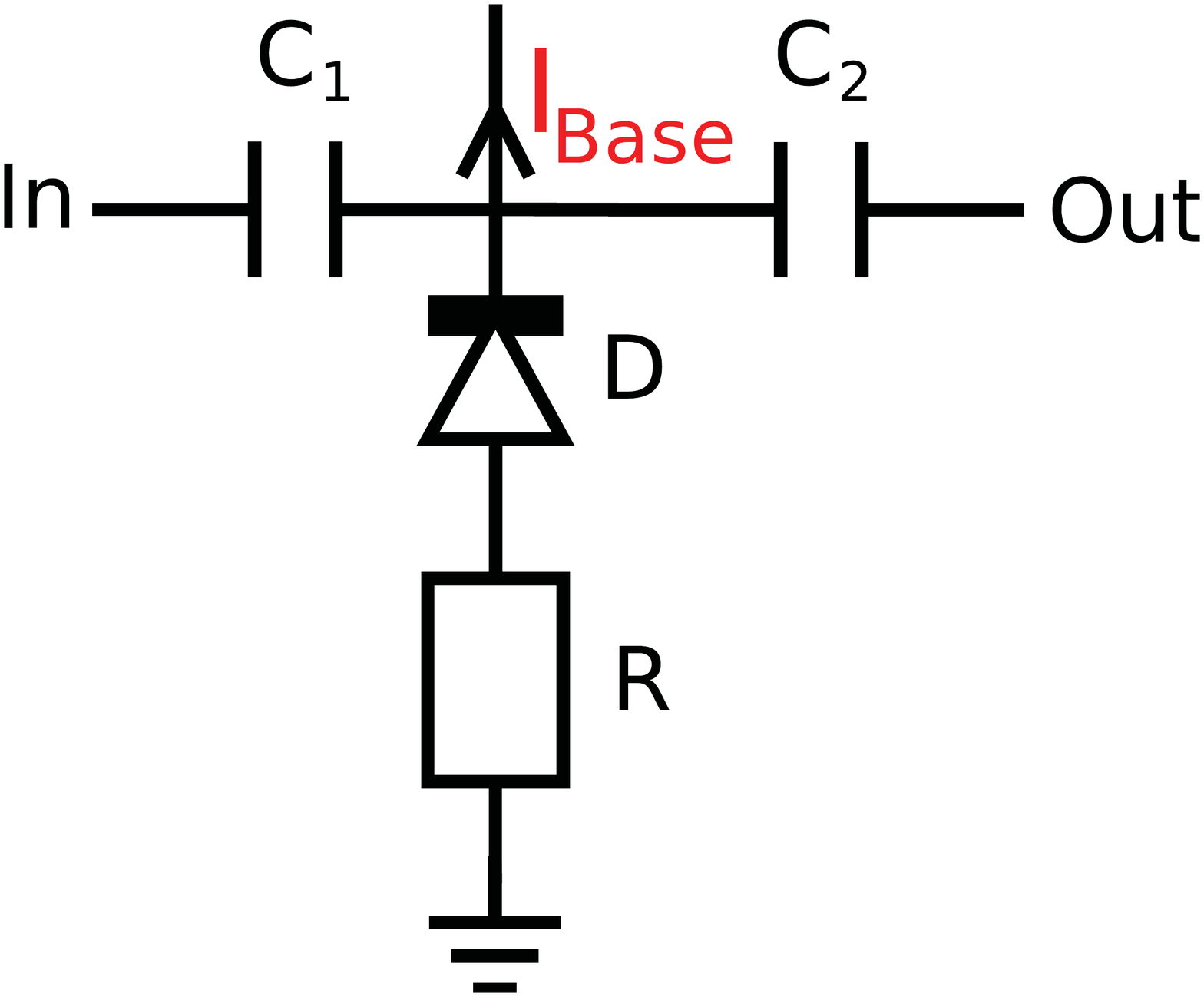}
		\caption{}
		\label{fig::blr_diode}
	\end{subfigure}
\vspace{-2mm}
	\begin{subfigure}[b]{0.5\textwidth}
		\centering
		\includegraphics[width=1.0\textwidth]{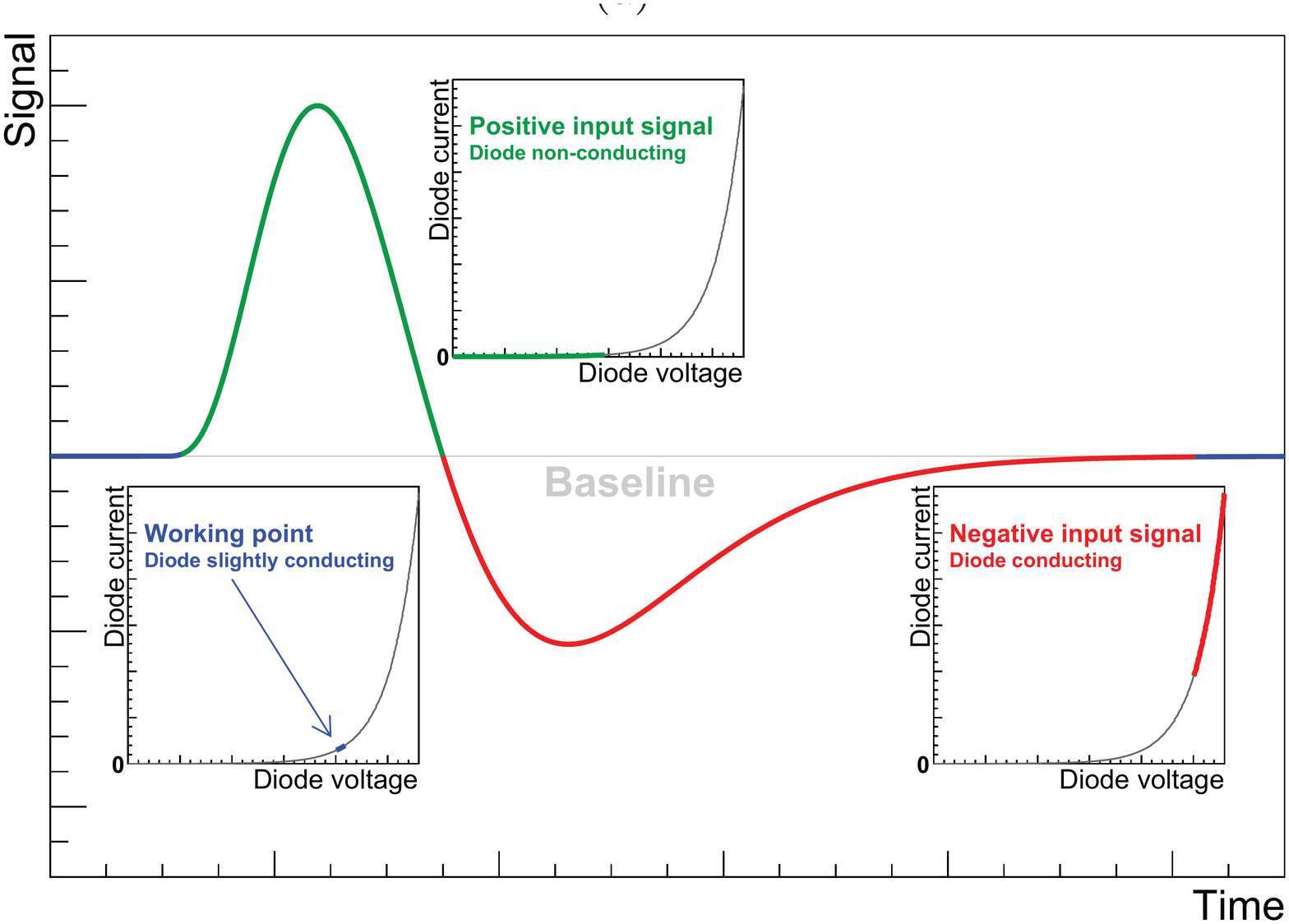}
		\caption{}
		\label{fig::blr_explain}
	\end{subfigure}
	\caption{(\subref{fig::blr_diode}) Circuit of a simple implementation of baseline restoration based on using a diode. (\subref{fig::blr_explain}) Working principle of the circuit (see text)~\cite{Pisa}.}
	\label{fig::blr}
\end{figure}

\section{Principle of a Baseline Restoration}

The pile-up effect can be suppressed using the concept of baseline restoration (BLR).
In \fig{fig::blr_diode} a simple baseline restoration circuit is shown which follows the bipolar shaper.
\fig{fig::blr_explain} explains the working principle using a diode with a working point set by the diode current $I_{Base}$~\cite{BLR}.
While for positive input signals the working point is shifted forwards in the non-conducting region of the diode characteristic leading to output signals unchanged compared to the input signal, negative signals are shorted with the conducting diode to ground leading to a fast restoration of the baseline.

\section{A Discrete Bipolar Shaping Circuit With Additional Baseline Restoration}

\begin{figure}[tbh]
	\centering
	\includegraphics[width=0.5\textwidth]{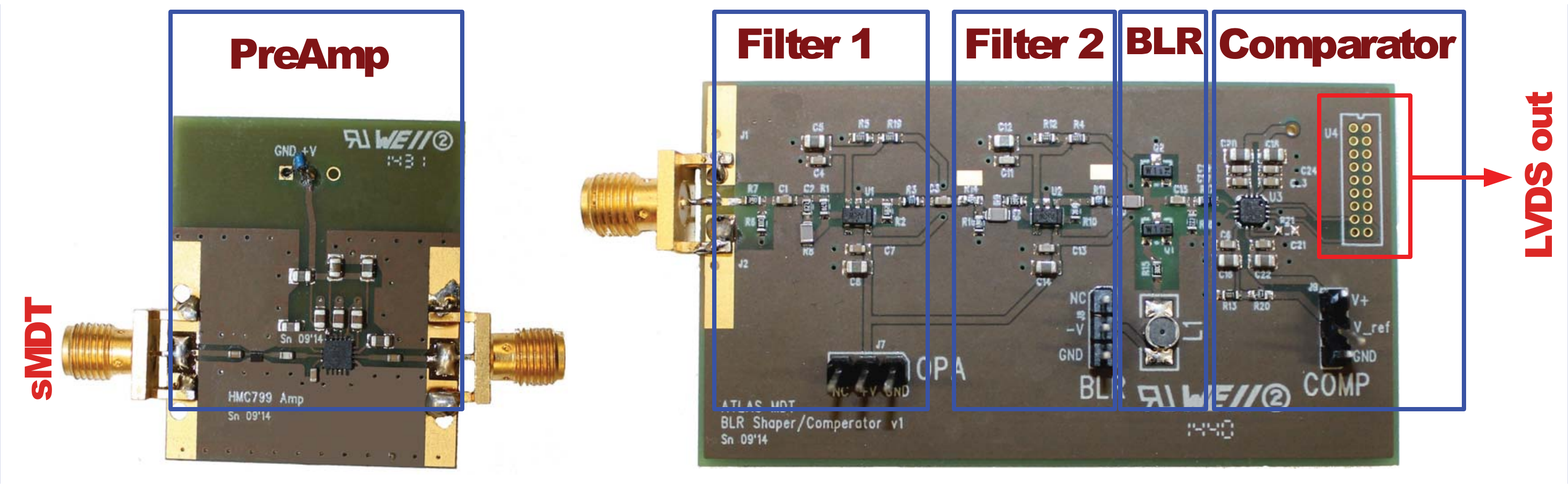}
	\vspace{-15mm}
	\caption{Discrete bipolar shaping circuit with baseline restorer (ASBC - Amplifier, Shaper, Baseline restorer, Comparator). A transimpedance amplifier (gain of 10000, 700~MHz bandwidth) is used as pre-amplifier, two filters and a baseline restorer for signal shaping.}
	\label{fig::ASBC}
\end{figure}

In order to investigate the effect of baseline restoration on sMDT signals and to perform studies for the design of a new ASD chip, a discrete amplifier and bipolar shaping circuit with optional BLR (ASBC - Amplifier, Shaper, Baseline restorer, Comparator) has been designed and built (see \fig{fig::ASBC}).
It consists of a transimpedance amplifier (gain of 10000, 700~MHz bandwidth) used as pre-amplifier, two filters for signal shaping, the BLR circuit shown in \fig{fig::blr_diode} and a comparator.
The peaking time of the ASBC is slightly shorter than of the ASD leading to suppression of time slewing effects\footnote{Jitter of the threshold crossing time due to the rise time of the shaped signal.}.
By setting the diode current $I_{Base}=0$ and $I_{Base}=90$~$\mu$A, the baseline restorer can be switched off and on, and, therefore, the impact of the baseline restoration can be studied in detail.
Due to the negligible dead time of the comparator (1.3~ns), the programmable dead time can be varied in data analysis leading to results with different programmable dead times. 

\begin{figure}[tbh]
	\centering
	\includegraphics[width=0.515\textwidth]{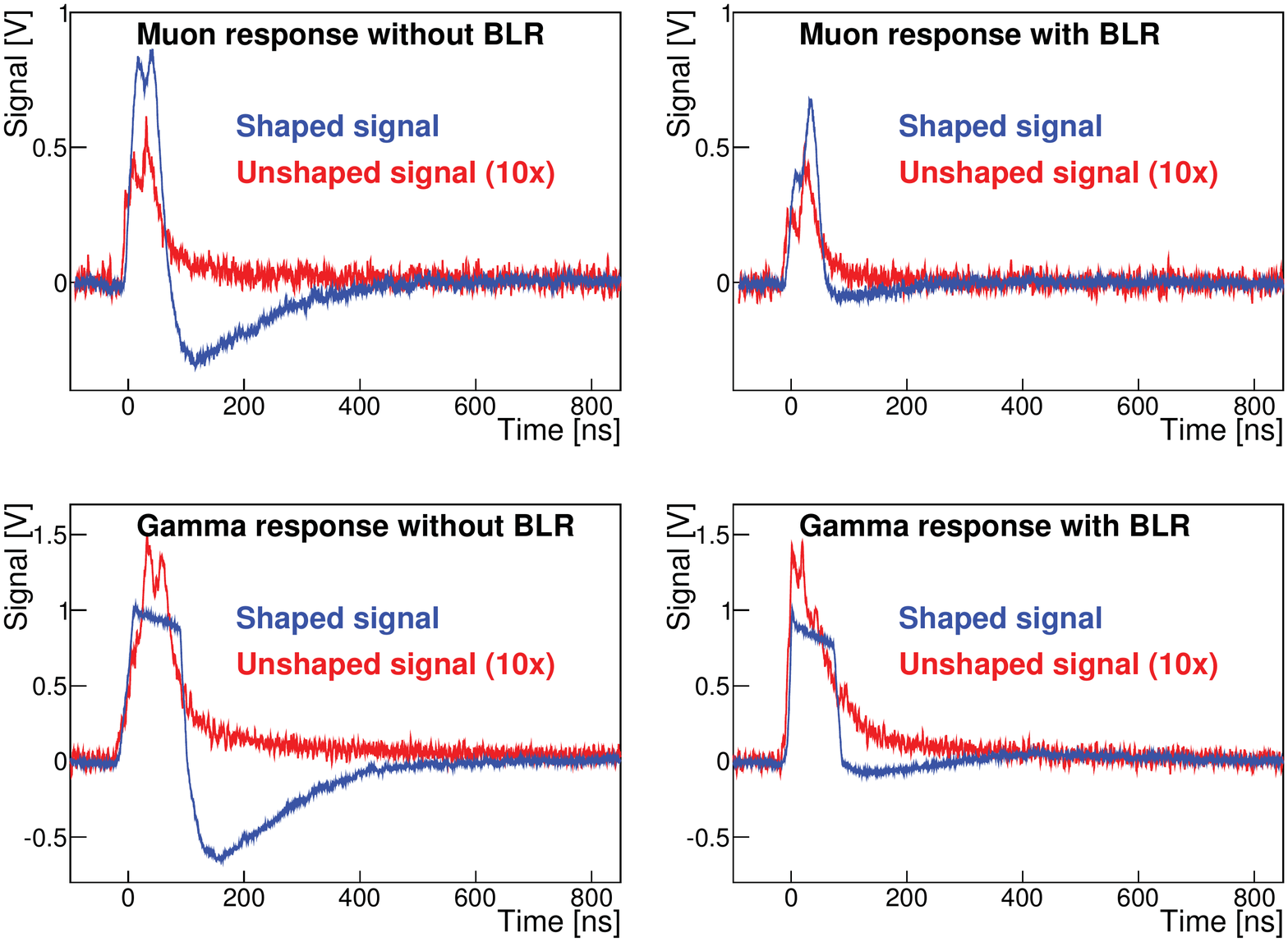}
	\caption{Unshaped (gain 10000) and bipolar shaped muon and $^{137}$Cs $\gamma$-signal of an sMDT tube without and with baseline restoration (200 MHz measurement bandwidth).}
	\label{fig::pulse}
\end{figure}

In \fig{fig::pulse} unshaped (amplifier gain of 10000) and bipolar shaped muon and $^{137}$Cs $\gamma$ signals in an sMDT tube with ASBC read-out without and with baseline restoration are shown.
The baseline restoration suppresses the undershoot and reduces the time until the baseline is restored by about a factor of two.

\section{Effect of BLR on Drift Tube Performance}

\begin{figure}[tbh]
	\centering
	\includegraphics[width=0.5\textwidth]{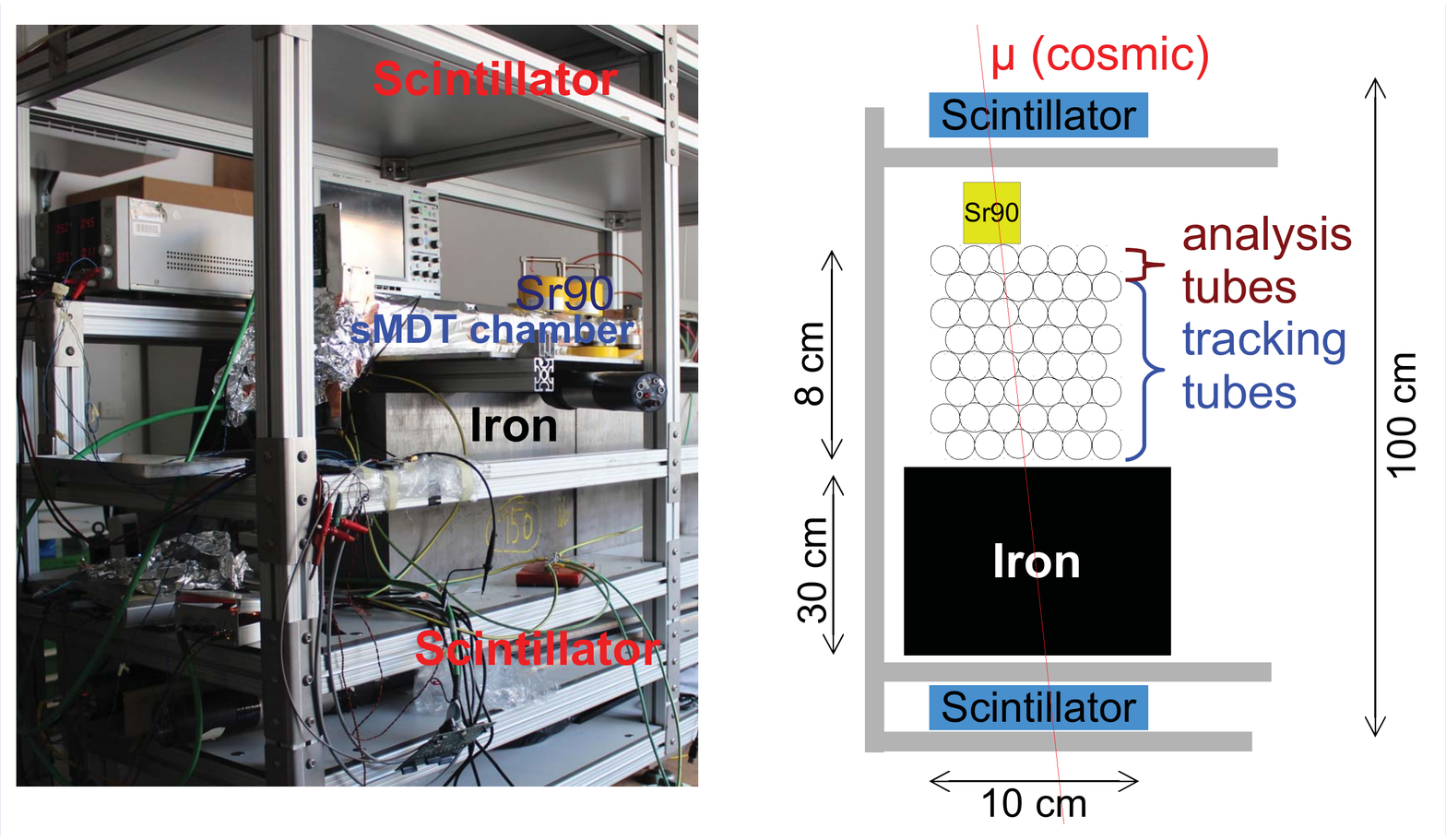}
	\caption{Experimental set-up. An sMDT chamber irradiated with a $^{90}$Sr source was used for the reconstruction of cosmic muon tracks. The scintillation detectors were used for triggering, the iron block for hardening of the cosmic muon momentum spectrum.}
	\label{fig::setup}
\end{figure}

An sMDT chamber has been irradiated with a 24~MBq $^{90}$Sr electron source (see~\fig{fig::setup}).
While seven out of eight sMDT layers were read-out with the standard ATLAS MDT read-out chip and used for muon track reconstruction, the eighth layer, so-called analysis tubes, was read-out with the ASBC.
Scintillation detectors were used as trigger and a 30~cm iron block for hardening of the muon spectrum.

\begin{figure}[h!]
	\centering
	\begin{subfigure}[b]{0.45\textwidth}
		\centering
		\includegraphics[width=1\textwidth]{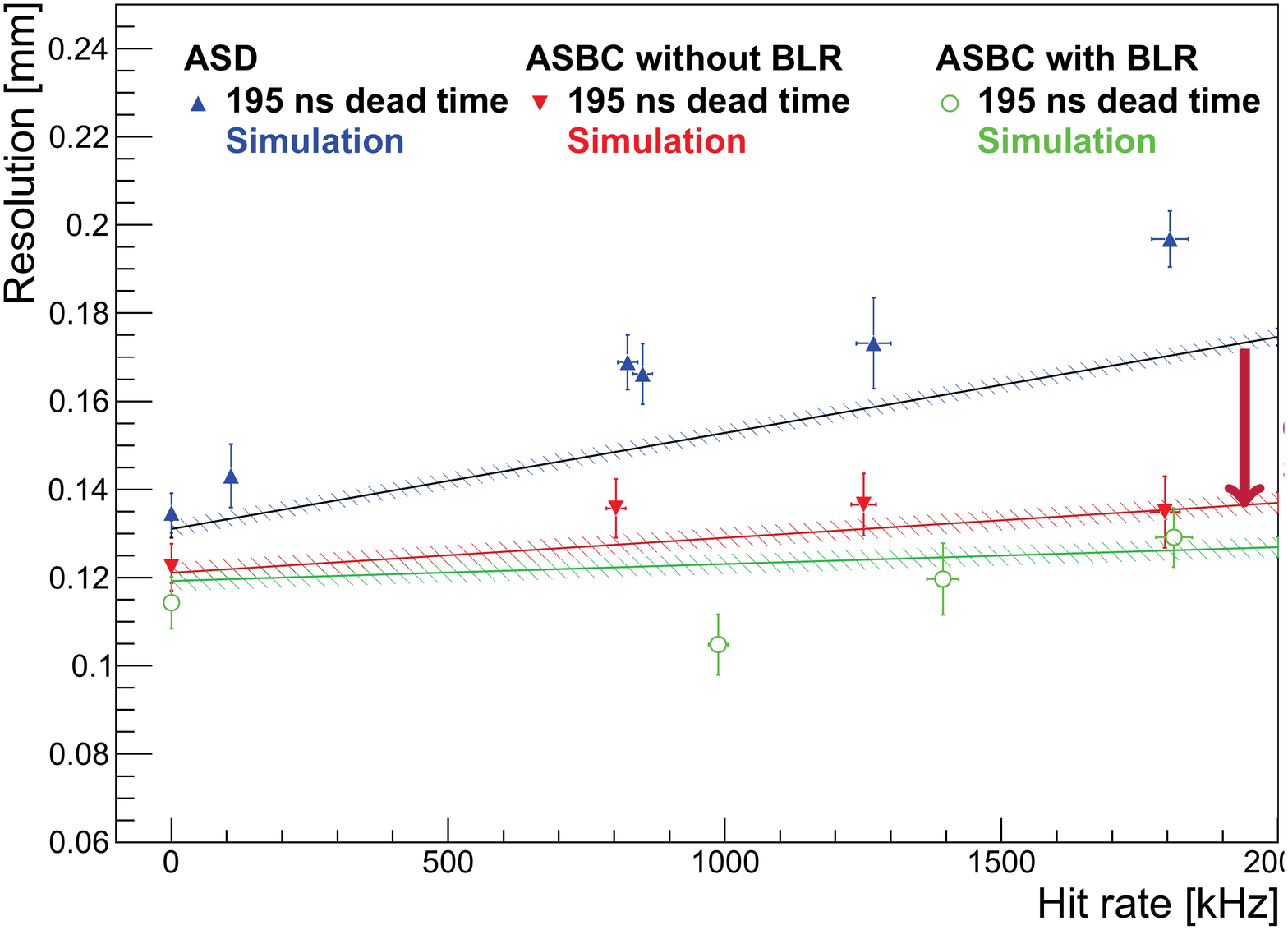}
		\begin{tikzpicture}[overlay],font=\small]
			\draw[color = purple,<-, line width=2pt] (3.4,3) to (3.4,3.95);
			\node[text width=1.8cm, color=purple] at (4.6,3.75) {Improved};
			\node[text width=1.8cm, color=purple] at (4.6,3.45) {due to};
			\node[text width=1.8cm, color=purple] at (4.6,3.15) {shaping};
		\end{tikzpicture}
		\caption{}
		\label{fig::res_elec}
	\end{subfigure}
	\begin{subfigure}[b]{0.45\textwidth}
		\centering
		\includegraphics[width=1.1\textwidth]{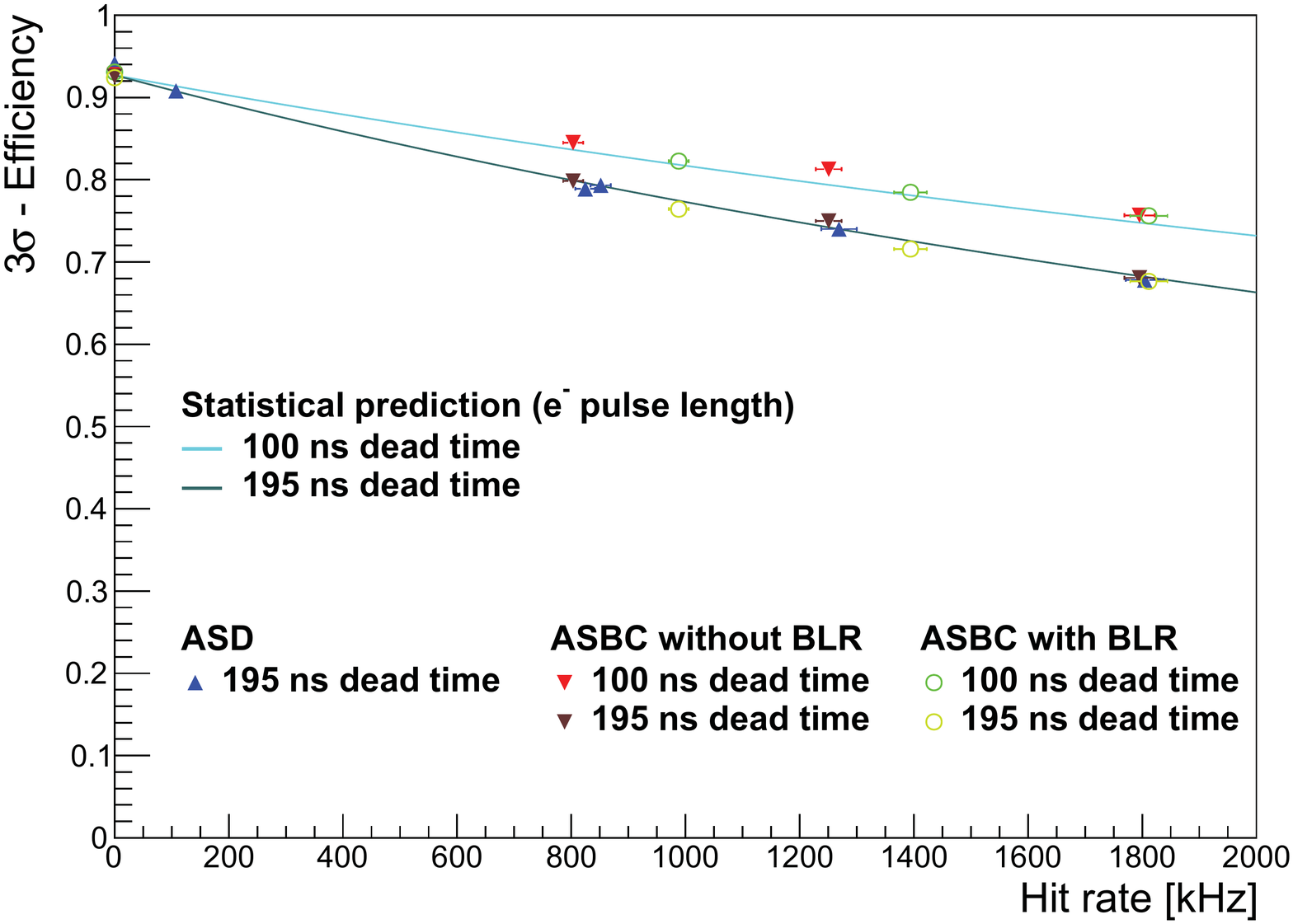}
		\caption{}
		\label{fig::eff_elec}
	\end{subfigure}
	\caption{Average single-tube spatial resolution and 3$\sigma$-efficiency of an sMDT tube measured with ASD and ASBC electronics with and without baseline restoration under $^{90}$Sr electron irradiation (see text).}
	\label{fig::reseff_elec}
\end{figure}

In \fig{fig::reseff_elec} the measured resolution and 3$\sigma$-efficiency\footnote{The probability for a hit to be measured on the extrapolated muon trajectory within three times the resolution.} as a function of the electron background hit rate with and without baseline restoration are shown in comparison with a reference measurement conducted with the ASD chip.
The resolution measurement is well described by a scattering parameter and SPICE based simulation.
The statistical prediction used to explain the measured 3$\sigma$-efficiency is based on the intrinsic read-out electronics dead time due to the pulse length of the background hits.
The results show that at high counting rates the measured sMDT single-tube resolution is improved substantially by the  shorter peaking time used for the ASBC  and can be further enhanced by using baseline restoration.

\begin{figure}[h!]
	\centering
	\begin{subfigure}[b]{0.45\textwidth}
		\centering
		\includegraphics[width=1.0\textwidth]{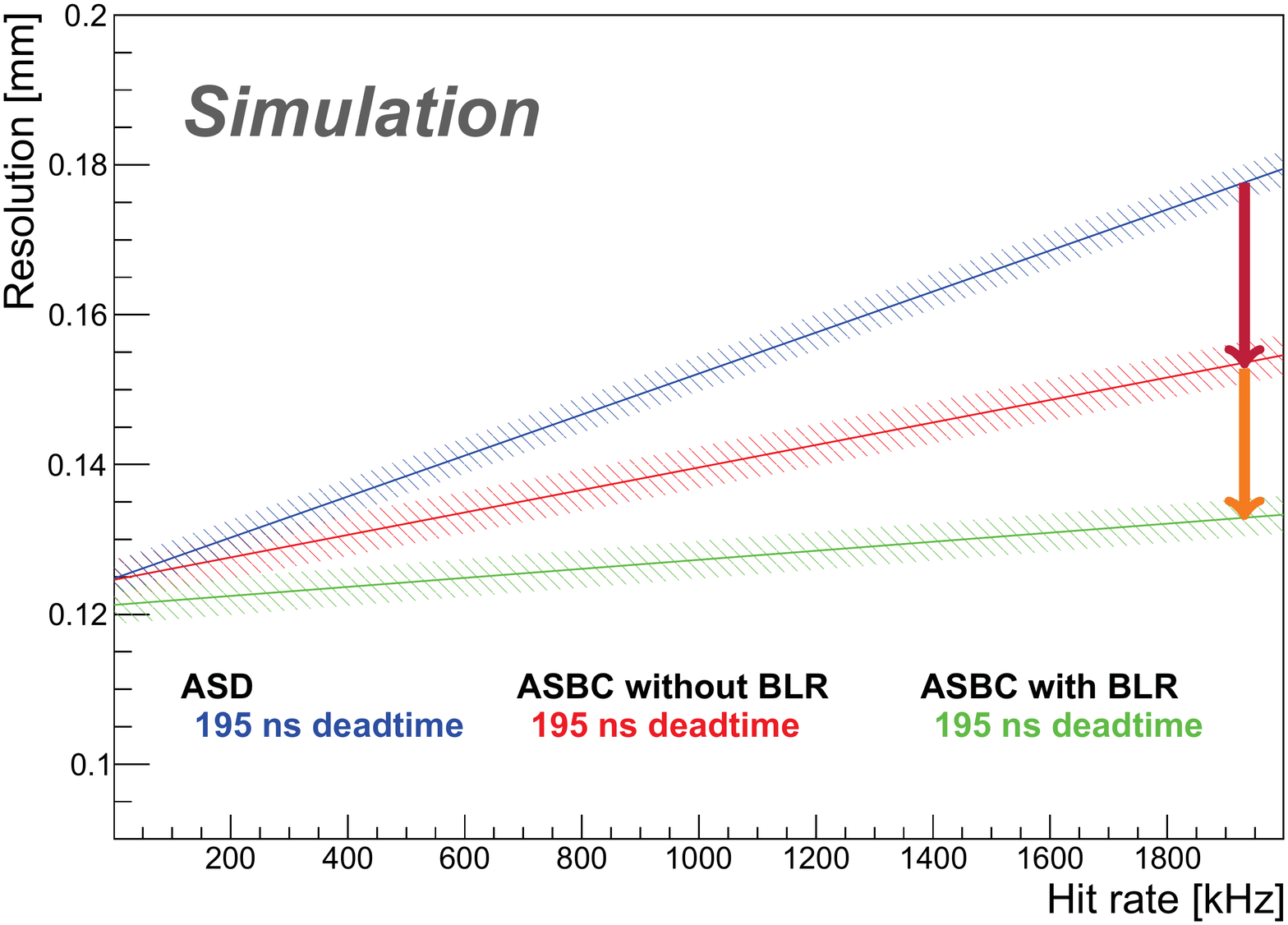}
		\begin{tikzpicture}[overlay],font=\small]
			\draw[color = purple,<-, line width=2pt] (3.6,3.9) to (3.6,4.95);
			\node[text width=1.8cm, color=purple] at (4.8,4.7) {Improved};
			\node[text width=1.8cm, color=purple] at (4.8,4.4) {due to};
			\node[text width=1.8cm, color=purple] at (4.8,4.1) {shaping};
			\draw[color = orange,<-, line width=2pt] (3.6,2.95) to (3.6,3.85);
			\node[text width=1.8cm, color=orange] at (4.8,3.7) {Improved};
			\node[text width=1.8cm, color=orange] at (4.8,3.4) {due to};
			\node[text width=1.8cm, color=orange] at (4.8,3.1) {BLR};
		\end{tikzpicture}
		\caption{}
		\label{fig::res_gamma}
	\end{subfigure}
	\begin{subfigure}[b]{0.45\textwidth}
		\centering
		\includegraphics[width=1.1\textwidth]{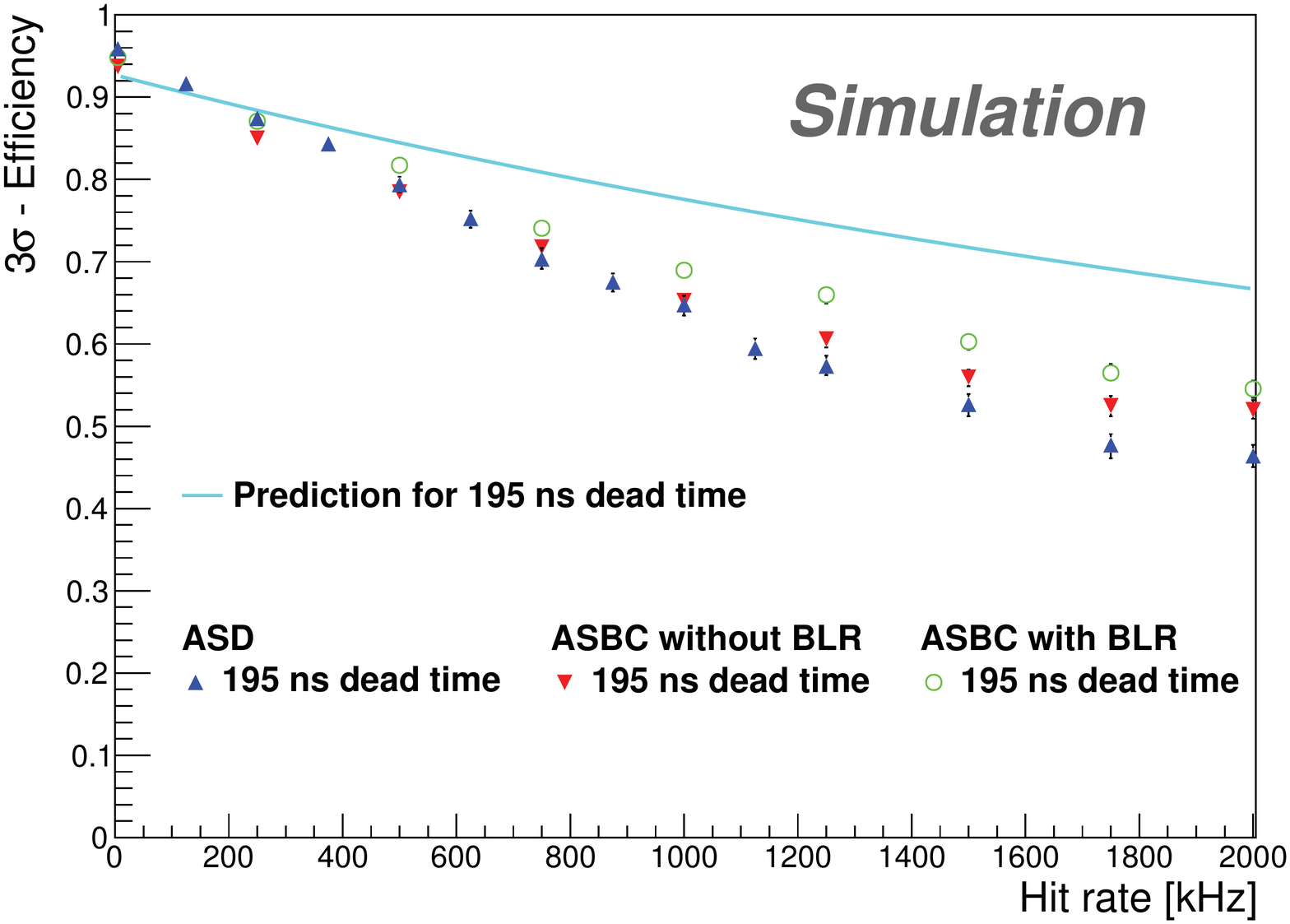}
		\caption{}
		\label{fig::eff_gamma}
	\end{subfigure}
	\caption{Average single-tube spatial resolution and 3$\sigma$-efficiency of an sMDT tube simulated for ASD and ASBC electronics with and without baseline restoration under $^{137}$Cs $\gamma$ irradiation (see text).}
	\label{fig::reseff_gamma}
\end{figure}

In order to gain a prediction for the impact of BLR at high counting rates due to $\gamma$-irradiation, the resolution and 3$\sigma$-efficiency have been simulated with $^{90}$Sr e$^{-}$ -pulses scaled to size of pulses from Compton electrons and taking gain drop due to space charge into account.
The results are shown in \fig{fig::reseff_gamma} and indicate that baseline restoration enhances the measurement resolution and efficiency substantially.

\section{Summary and Conclusion}

The performance of sMDT tubes at high background counting rates can be further improved for the use at HL-LHC and future hadron colliders by improving the rate capability of the read-out electronics using baseline restoration in addition to the bipolar shaping.
A substantial improvement of the drift tube spatial resolution at high counting rates has been demonstrated in measurements and simulations with a discrete ASD circuit with additional baseline restoration functionality.

\end{document}